\documentclass[conference]{IEEEtran}
\IEEEoverridecommandlockouts
\usepackage{cite}
\usepackage{amsmath,amssymb,amsfonts}
\usepackage{algorithmic}
\usepackage{graphicx}
\usepackage{textcomp}
\usepackage{xcolor}
\usepackage{hyperref}
\usepackage{multirow}
\let\OLDthebibliography\thebibliography
\renewcommand \thebibliography[1]{
  \OLDthebibliography{#1}
  \setlength{\parskip}{0pt}
  \setlength{\itemsep}{0pt plus 0.3ex}
}
\def\BibTeX{{\rm B\kern-.05em{\sc i\kern-.025em b}\kern-.08em
    T\kern-.1667em\lower.7ex\hbox{E}\kern-.125emX}}
    
\begin{document}

\title{Picking watermarks from noise (PWFN): an improved robust watermarking model against intensive distortions\\
\thanks{$\ast$corresponding author}
\thanks{This work was supported in part by the Natural Science Foundation of China under Grant 61972169, in part
by the National key research and development program
of China(2019QY(Y)0202, 2022YFB2601802), in part by
the Major Scientific and Technological Project of Hubei
Province(2022BAA046, 2022BAA042), in part by the Research Programme on Applied Fundamentals and Frontier
Technologies of Wuhan(2020010601012182) and the Knowledge Innovation Program of Wuhan-Basic Research, in part
by China Postdoctoral Science Foundation 2022M711251.}
}

\author{\IEEEauthorblockN{ Sijing Xie}
\IEEEauthorblockA{\textit{school of Computer Science and Technology} \\
\textit{Huazhong University of Science and Technology}\\
Wuhan, China \\
xiesijing@hust.edu.cn}
\and
\IEEEauthorblockN{ Chengxin zhao}
\IEEEauthorblockA{\textit{school of Computer Science and Technology} \\
\textit{Huazhong University of Science and Technology}\\
Wuhan, China \\
zhaochengxin@hust.edu.cn}
\and
\IEEEauthorblockN{ Nan Sun}
\IEEEauthorblockA{\textit{school of Computer Science and Technology} \\
\textit{Huazhong University of Science and Technology}\\
Wuhan, China \\
sunnan@hust.edu.cn}
\and
\IEEEauthorblockN{Wei Li}
\IEEEauthorblockA{\textit{School of Software} \\
\textit{Nanchang University}\\
Nanchang, China \\
weili.cs@ncu.edu.cn}
\and
\IEEEauthorblockN{Hefei Ling$^{\ast}$}
\IEEEauthorblockA{\textit{school of Computer Science and Technology} \\
\textit{Huazhong University of Science and Technology}\\
Wuhan, China \\
lhefei@hust.edu.cn}
\and
}

\maketitle

\begin{abstract}
Digital watermarking is the process of embedding secret information by altering images in an undetectable way to the human eye. To increase the robustness of the model, many deep learning-based watermarking methods use the encoder-noise-decoder architecture by adding different noises to the noise layer. The decoder then extracts the watermarked information from the distorted image. However, this method can only resist weak noise attacks. To improve the robustness of the decoder against stronger noise, this paper proposes to introduce a denoise module between the noise layer and the decoder. The module aims to reduce noise and recover some of the information lost caused by distortion. Additionally, the paper introduces the SE module to fuse the watermarking information pixel-wise and channel dimensions-wise, improving the encoder's efficiency. Experimental results show that our proposed method is comparable to existing models and outperforms state-of-the-art under different noise intensities. In addition, ablation experiments show the superiority of our proposed module.
\end{abstract}

\begin{IEEEkeywords}
neural network, robust watermarking, image denoise
\end{IEEEkeywords}

\section{Introduction}
Digital watermarking is a crucial aspect of information security, serving as the primary method for copyright protection, leakage traceability, and proactive forensics\cite{Zhang_Lin_Benz_Chen_Zhang_Kweon_2021} \cite{Wang_Byrnes_Wang_Sun_Ma_Chen_Wu_Xue_2021}. This technology allows for the embedding of specified information in images, text, and video carriers, which can be extracted when necessary to identify the ownership of the work. Watermarked images are frequently subject to intentional or unintentional attacks during dissemination, which can result in image distortion. These distorted images may cause the failure of watermark extraction.
There are digital watermarking methods with higher robustness that can solve this problem. Some existing methods have already achieved a certain degree of robustness. However, these methods are only resistant to certain weak noise attacks and may not meet the requirements for practical use. 

To improve the robustness of watermarking algorithms, we introduce the denoise module to the existing encoder-noise-decoder framework, which we call PWFN. The distorted embedded image is initially transmitted to the denoising module for denoising prior to decoding. 
The denoising module serves to attenuate noise interference to the image, thereby facilitating the subsequent decoding operation. From this idea, we innovatively introduce the denoise module into the watermarking task and achieve higher decoding accuracy by using the recovered image for watermark extraction.

% This weakens the damage caused by noise to the watermark information and improves the methods's robustness. 

In addition, the encoder makes use of the SE module\cite{SE} to improve the coupling of the watermark information with the original image, resulting in improved robustness and visual quality.
\begin{figure}[t]\label{img:enbeded image result}
  \centering
  \includegraphics[width=0.5\textwidth]{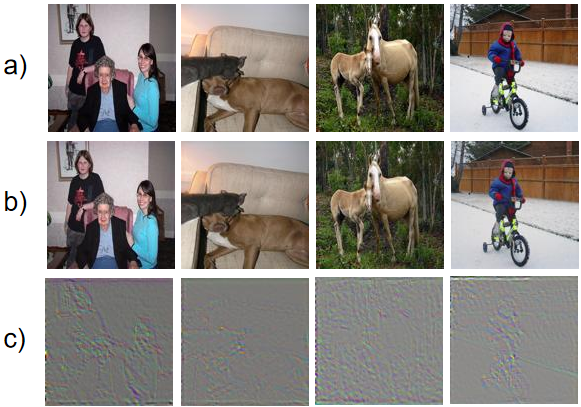}
  \caption{Our approach embeds the given message into the image with enhanced robustness and better visual quality. (a)original images,(b)encoded images,(c) residual images.}
\end{figure}
By optimizing the encoder embedding the message methods and introducing a new denoising module into the network, we aim to reduce the impact of noise on the watermark image, thereby further improving decoding accuracy.

% Extensive experiments validate the effectiveness of our approach. Further ablation experiments underscore the robustness of our method. We achieve enhanced robustness of the watermark by training a module specifically designed to restore the watermark.

Our primary contributions are as follows:
\begin{itemize}
    \item The watermark embedding framework was enhanced by incorporating a denoise module into the network. This module restores degraded images, transforming the watermark robustness task into a denoise task.
    \item To improve the coupling of the container image and watermark information, we suggest a coding approach that combines the pixel and channel dimensions.
    \item Compared to the existing SOTA methods, our approach achieves comparable results under low noise intensity and surpasses SOTA models as noise intensity increases, demonstrating superior visual quality.
\end{itemize}

\section{Related work}
\subsection{Deep learning Digital Watermarking}
% Traditional methods, deep learning methods (END ,MBRS) 
%  Digital watermarking algorithms can be mainly divided into traditional digital watermarking methods and digital watermark embedding methods based on deep learning. Traditional digital watermarking usually chooses to embed the corresponding watermark information in the null domain or transform domain, such as the least significant bit watermark embedding method. Transform domain embedding methods mainly include DWT \cite{DWT}, DCT \cite{DCT} and Haar \cite{haar} decomposition, etc., which resist specific noise attacks by changing the low and medium frequency coefficients in the frequency domain. Although these traditional methods can realize watermark embedding, most of them are very complex and have low robustness.	

Deep learning has shown great strength in the field of watermarking. Zhang et al. \cite{hidden} proposed an end-to-end deep learning
framework for watermark embedding. The model improves the robustness by adding different types of noise to the layers through a noise layer in the encoder and a noise layer in the decoder. MBRS  \cite{MBRS_Jia_Fang_Zhang_2021} proposes a novel method by alternately training real JPEG, simulated JPEG, and distortion-free images in a mini-batch method to improve the real JPEG robustness. Hao et al. \cite{Hao_Feng_Zhang_2020} propose to use high-pass filtering before the discriminator input to force the model to embed the watermark information in the low and mid-frequency regions of the image to improve the robustness of the model. Arwgan \cite{arwgan} proposes to add an attention mechanism (AM) 
to make the model pay attention to the robustness area of an image and a feature fuse module (FFM) to the encoder to fuse the watermark information and container image feature to improve the robustness of the model future. 
    
These deep learning methods are all based on END architecture model design, and the core of their processing is to modify the encoder and noiser layers to obtain a more robust watermarking model, ignoring the design that can be done between the distorted image and the decoder.
\subsection{Image Denoising}
% Traditional methods, deep learning based methods DCNN, UNET, KHIT.
Image denoising and restoration is a very popular task in the field of computer vision and has attracted the attention of many researchers with the aim of restoring images after distortion. Traditional image denoising methods use the local similarity of the image to perform the restoration operation. BM3d \cite{dabov2009bm3d} proposes a traditional image denoising algorithm based on block matching, which reduces the noise level by performing a chunking operation on the image using the similarity between the chunks of the image. NLMeans \cite{nlmbs} reduces the noise level by using the non-local similarity in the graph.
    
DnCNN \cite{DncNN}, as a classical deep denoising model, achieves good denoising results by using CNN to learn the noise features of an image. Unet network is trained by successive downsampling and upsampling to obtain clean images. Noise2Noise \cite{noiser2noiser} proposes a self-supervised image denoising method that can be trained without clean image labels and learns the denoising network only by using pairs of noisy images for training to learn the denoising network. Pix2Pix \cite{pix2pix}introduces adversarial generative networks to the task of image denoising by training generators and discriminators to learn the denoising mapping of an image. Image denoising can reduce some of the distortion caused by noise.

\section{Methods}

\begin{figure}\label{img: mixedup}
  \centering
  \includegraphics[width=0.45\textwidth]{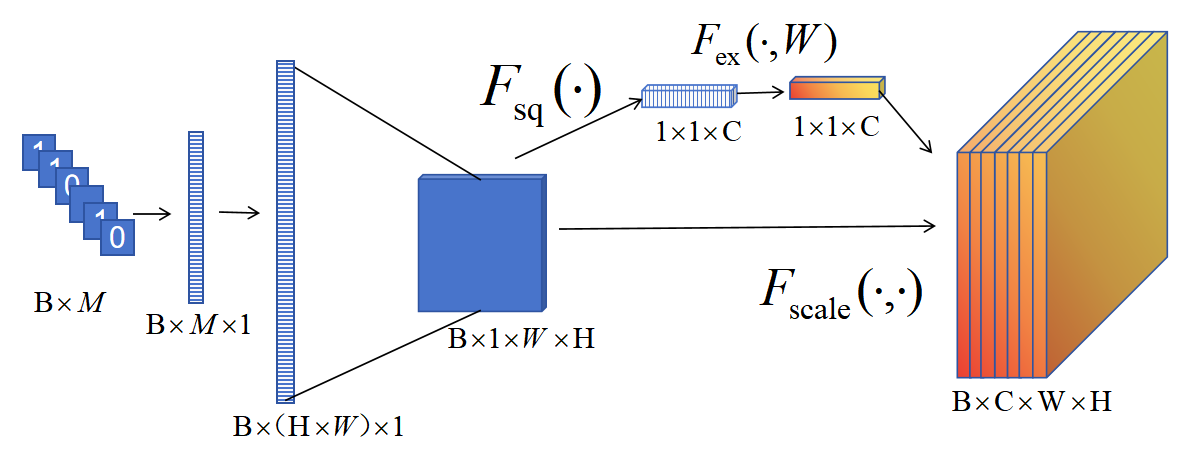}
  \caption{Mixes the pixel-wise and channel-wise information of the watermark, where $F_{sq}(\cdot)$, $F_{ex}(\cdot)$ and $F_{scale}(\cdot)$ means the squeezing, excitation, and scaling operation respectively.}
\end{figure}

\begin{figure*}[ht!]\label{img:network pipline}
  \centering
  \includegraphics[width=0.8\textwidth]{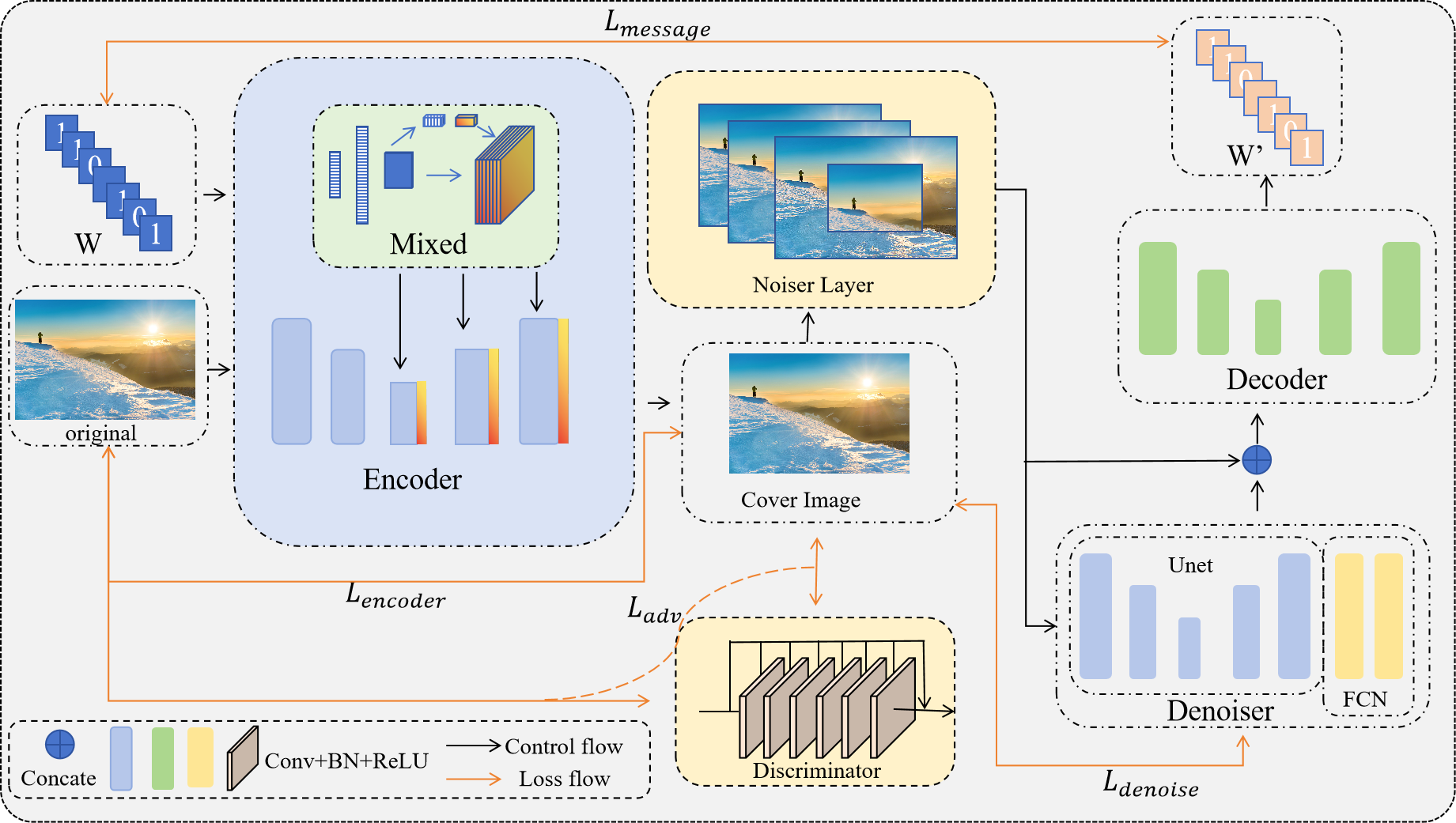}
  \caption{Overview of the proposal network framework. The encoder generates the encoded image using the image and watermark information. The encoded image is then subjected to common noise processing to create a distorted image that resists noise. Then, the denoiser module reduces the distortion caused by the noise in the coded image. Finally, the decoder extracts the embedded watermark information.}
\end{figure*}

%  Introduction to the structure of ENRD structure
  In this section we propose an improved robust watermarking model against intensive distortions named \textbf{p}king \textbf{w}atermarks \textbf{f}rom \textbf{n}oise (PWFN), which innovatively introduces a SE-encoder and a denoiser module, as shown in figure \ref{img:network pipline}. Each module of the method is depicted in more detail below.
% Figuer 2. (Overall structure of the network)

% Detailed description of the E-structure and R-structure
% The E-structure and the R-structure are divided into encoder, decoder, noise layer, denoise layer, and discriminator.
% ARES: On Adversarial Robustness Enhancement for Image Steganographic Cost Learning sections attention mechianism
\subsection{Encoder}
   The role of the encoder is to embed the watermark information $W^{'}$ into the original image. In order to obtain a high visual quality carrier image, the generated result of the network is constrained by designing the loss function $L_{encoder}$. The equation is defined below:
   \begin{equation}
       L_{enocder} = MSE(I_{en},I_{ori})
   \end{equation}
   Where $MSE(\cdot)$ is the mean square error, $I_{en}$ is the carrier image, and $I_{ori}$ is the original image.
In order to further improve the ability of the encoder to embed the container image, we improve the original encoder structure and propose a hybrid embedding method of pixel-wise and channel-wise, which can better realize the coupling of carrier image and watermark information. It can reduce the embedding of redundant watermark information and improve the visual quality of the network. Its design is shown in Fig \ref{img: mixedup}.
\begin{equation}
    W^{'} = SE(Linear(W))
\end{equation}
For pixel-wise, linear transform is used to realize the dispersion process of watermark information $W \in R^{B\times M \times 1 }$, which will generate a 16$\times $16 vector $ W^{'} \in R ^{B\times1\times H \times W}$, and then for this feature SE\cite{SE} module is used to process the watermark information in the channel dimension. Where $W$ is the Message length, $B$ is the batch size and $H$ and $W$ mean the height and width of an image respectively. Then, the watermark information and carrier image coupling are realized by concatenating to the carrier image. 
 % Details are given in the \textbf{supporting material}.
The final encoding process is expressed as follows. Where $E(\cdot)$ represents the encoding process.
\begin{equation}
    I_{en} = E(I_{ori},W^{'})
\end{equation}
% (Example image showing the proposed easing feature embedding method, and the corresponding pixel embedding method)
% (Description of Eq.)

% We use linear to accomplish the dispersion process of the watermarked information and introduce SE network to better incorporate the watermarked features in the channel dimension. We also make flexible use of jump-joins to fully couple the image and the watermarked features to obtain a more robust carrier image.The SE-structure proposes a method for comparing the corresponding pixel and channel wise methods, and suggests that this method can be used to better enable the encoder to embed the corresponding message information, and to improve the robustness of the embedded watermarked information. By using the SE module, the compression and topology operations of the channel dimension are realized to improve the feature coupling between the watermark information and the image during the encoder process.
\begin{figure*}[!ht]\label{img:compare with arwgan}
  \centering
  \includegraphics[width=0.8\textwidth]{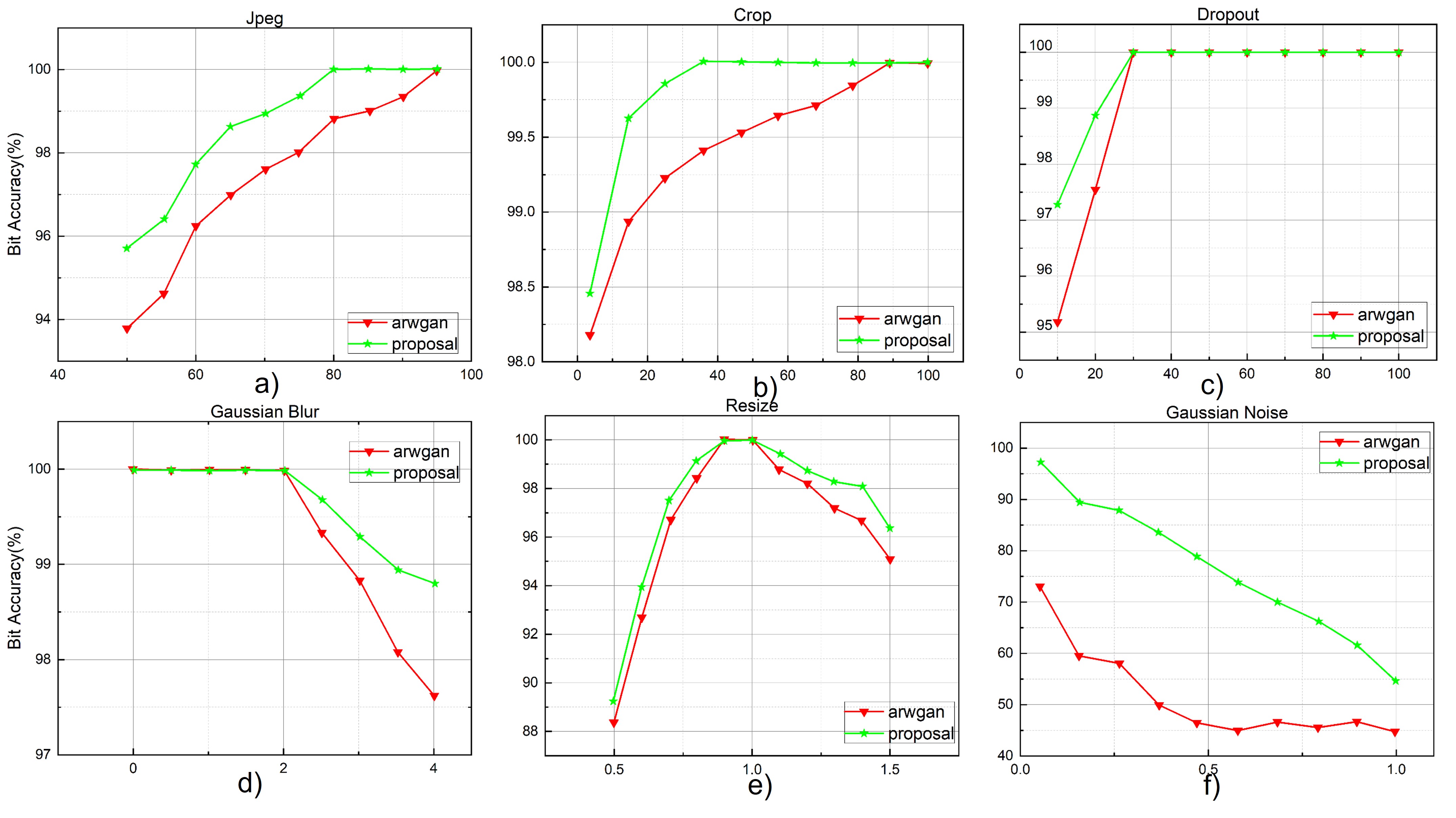}
  \caption{Robustness comparison with baseline network under different noise levels.(a)Jpeg,(b)Crop,(c)Dropout,(d)Gaussian Blur,(e)Resize, (f)Gaussian Noise.}
\end{figure*}

\begin{table*}[!ht]
    \centering
    \caption{Comparsion of different models on COCO \cite{COCO}}
    \resizebox{0.95\linewidth}{!}{
    \begin{tabular}{c|cc|ccccc|c}
    \hline
        \multirow{2}{*}{Model} & \multicolumn{2}{c|}{Invisibility}  & ~ & ~ & Robustness[\%] & ~ & ~ & \multirow{2}{*}{Average} \\ \cline{2-8}
        ~ & PSNR & SSIM & JPEG(50) & Cropout(30\%) & Dropout(30\%) & Crop(3.5\%) & Gaussian Blur(2.0) & ~ \\ \hline
        ReDMark \cite{ReDMArk} & 35.93  & 0.97  & 74.60  & 92.50  & 92.00  & 100.00  & 50.00  & 81.82  \\ \hline
        DA  \cite{DA}& 33.70  & --- & 81.70  & --- & 97.90  & 93.50  & 60.00  & 83.28  \\ \hline
        TSDL \cite{TSDL} & 33.50  & --- & 76.20  & 97.30  & 97.40  & 89.00  & 98.60  & 91.70  \\ \hline
        MBRS \cite{MBRS_Jia_Fang_Zhang_2021}& 35.84  & 0.89  & 91.97  & 99.58  & 99.96  & 92.68  & \textbf{100.00}  & 96.83  \\ \hline
        SSLW \cite{SSLW}& 34.00  & 0.87  & 83.01  & 79.66  & 88.11  & 50.73  & 98.96  & 80.09  \\ \hline
        ARWGAN \cite{arwgan}& 35.87  & 0.97  & 93.89  & 99.82  & 100.00  & 98.17  & 99.99  & 98.37 \\ \hline
        Proposal & \textbf{36.34}  & \textbf{0.98}  & \textbf{95.85}  & \textbf{99.93}  & \textbf{100.00}  & \textbf{98.45} & 99.99
  & \textbf{98.84}  \\ \hline
    \end{tabular}
    }
    \label{tb: result in coco}
\end{table*}
\subsection{Noiser Layer}
 The noiser layer is usually designed between the encoder and decoder modules. By applying noise to the container image carrying the watermark information, the encoder can be motivated to embed the information in a more stable region of the image and thus become somewhat robust to noise.
Crop, Dropout, Gaussian Blur, Resize, JPEG, and other noises have been added to the noise layer to simulate the distortions that would occur if the image were actually used. To simulate JPEG noise, we use the microscopic JPEG method proposed by MBRS\cite{MBRS_Jia_Fang_Zhang_2021}. This is necessary to avoid the problem of gradient vanishing during backpropagation caused by quantization. This process can be described by the following equation \ref{eq:noiser}. Where $Noiser(\cdot)$ means Noiser operation.
\begin{equation} \label{eq:noiser}
    I_{no} = Noiser(I_{en})
\end{equation}
\subsection{Denosier}
The denoiser module receives a noisy processed image as input. Its purpose is to reduce the interference of noise on the watermarked image, allowing for the recovery of watermark information that may have been damaged by the noise. 
Given the limitations of the denoising module in reconstructing the image, it is more probable that it will reconstruct the low-frequency regions. Consequently, under the constraints of the loss function, the watermarked signal will tend to be embedded in the low-frequency part of the image. Additionally, the watermarked signals in the low-frequency part possess greater robustness\cite{Hao_Feng_Zhang_2020}, which explains the effectiveness of our proposed module.
Specifically, we use a Unet \cite{unet} network and an FCN \cite{FCN} network to generate the residuals of the original image and the encoded image. The structure is shown in Figure \ref{img:network pipline} description.It could be described as below.
% This is expressed by the following equation \ref{eq:denosier}.
\begin{equation}\label{eq:denosier}
    L_{denoiser} = \frac{1}{2N} \sum_{i=1}^{N}\Vert R(I_{noiser}, \Theta) - (I_{en} - I_{ori})\Vert_F^{2}
\end{equation} 
Where $\Theta$ is the trainable parameters of the denoise module, $R(\cdot)$ is the residual between the original Image($I_{ori}$) and the embedded image($I_{en}$), $\Vert \cdot \Vert_F^{2}$ means the $L_2$ normal.
Compared to existing methods, our method denoise the distorted image before the watermark information is extracted, which reduces the noise interference on the watermark extraction and can retrieve part of the watermark information, which leads to higher robustness. Theoretically, the better the performance of the denoising network, the more effective it will be in improving the decoding performance of the network.

\subsection{Decoder}
The decoder comes to recover the watermark information embedded in the encoder from the image after denoising. The decoder receives the distorted image through a skip connection and image stitching from the denoising module. It can be described as the following equation \ref{eq:decoder}. Where $I_{re}$ and $I_{noiser}$  represent the output of the denoising module and the image after noise distortion, respectively.
A hard-coded layer is used here equation \ref{eq: hardcoder}.
\begin{equation}\label{eq:decoder}
    W_{out} = Decoder(concate(I_{re},I_{noiser}))
\end{equation}
\begin{equation}\label{eq: hardcoder}
    W_{out} = Hard\_Threshold(S, 0.5)
\end{equation}
Where values greater than 0.5 are considered as 1 and values less than 0.5 are considered as 0. The loss function is calculated using the following equation \ref{eq:loss decoder}.
\begin{equation} \label{eq:loss decoder}
    L_{decoder} = \frac{\Vert W_{in} - W_{out} \Vert_{2}^{2}}{L}
\end{equation}
Where  $W_{in}$ and $W_{out}$ are the original watermark information and the extracted watermark information, respectively.
\subsection{Discrimator}
To avoid image artifacts that tend to occur in watermarked images and to obtain higher visual quality.We choose PatchGAN \cite{isola2017image}as the discriminator since it induces the encoded block to retain more details of the original image and reduce artifacts in the encoded image.%A discriminator is introduced to make the model generate more realistic watermarked images.% It is used to discriminate between the image generated by the  encoder, labeled 0, and the original input carrier image, labeled 1. The distribution of the watermark in the image is further adjusted through adversarial training to obtain a watermarked image with better visual quality. The loss function of the discriminator can be described by the following equation \ref{eq:dis}.
\begin{table*}[ht]
    \centering
    \caption{Comparison of different methods on unseen datasets}
    \begin{tabular}{c|ccc|ccc|ccc}
    \hline
        \multirow{2}{*}{Methods} & ~ & IMAGENET \cite{IMageNet} & ~ & ~ & DIV2K\cite{div2k}  & ~ & ~ & VOC2012\cite{pascal-voc-2012} & ~ \\ \cline{2-10}
        ~ & PSNR & SSIM & BAR & PSNR & SSIM & BAR & PSNR & SSIM & BAR \\ \hline
        HiDDeN \cite{hidden} & 33.26 & 0.93322 & 91.57 & 33.53 & 0.9272 & 91.93 & 33.35 & 0.8873 & 91.33 \\ \hline
        MBRS \cite{MBRS_Jia_Fang_Zhang_2021} & 36.44 & 0.8919 & 92.64 & 36.19 & 0.9009 & 96.99 & 35.84 & 0.8899 & 96.62 \\ \hline
        SSLW \cite{SSLW} & 33.5 & 0.8412 & 79.13 & 34.2 & 0.8513 & 79.79 & 34.12 & 0.8725 & 78.05 \\ \hline
        ARWGAN \cite{arwgan} & 36.66 & \textbf{0.9685} & 98.8 & 36.39 & \textbf{0.9623} & 98.72 & 36.83 & 0.9698 & 98.89 \\ \hline
        Proposal & \textbf{36.73} & 0.9672 & \textbf{99.12} & \textbf{36.43} & 0.9618 & \textbf{98.73} & \textbf{37.43} & \textbf{0.9712} & \textbf{98.91} \\
     \hline
    \end{tabular}
    \label{tb:generlization_compare}
\end{table*}
\begin{equation}\label{eq:dis}
    L_{discrimator} = [log(D(\theta, I_{ori})) +log(1 -D(\theta, I_{en}))]
\end{equation}
Where $\theta$ is the trainable parameters of the discriminator $D$, $I_{ori}$ and $I_{en}$ denoted original and encoded images respectively. 

\subsection{ Total Loss}
% Loss Function.
The overall loss function of the network can be described by the following equation \ref{eq:loss function}.
\begin{equation} \label{eq:loss function}
    L = \lambda_1 L_{enocoder} + \lambda_2L_{decoder}\\ + \lambda_3 L_{discrimator} + \lambda_4 L_{denoise}
\end{equation}
Where $\lambda_1, \lambda_2,\lambda_3, \lambda_4$ represent the hyperparameters respectively.

% Compared to existing methods, our method denoises the distorted image before the watermark information is extracted, which reduces the noise interference on the watermark extraction and can retrieve part of the watermark information, which leads to higher robustness. 
The network uses an end-to-end training approach to simultaneously optimize the encoder, discriminator, decoder, and denoiser modules in the network.

\section{Experiment and Results}\label{experiment and results}
As our model is based on ARWGAN\cite{arwgan}, we compared it at different noise intensities, as shown in Fig\ref{img:compare with arwgan}. This clearly demonstrates the superiority of our proposed method.
% As our baseline model is ARWGAN \cite{arwgan}, we compare different levels of noise with it to demonstrate the efficiency of our model, as shown in Figure \ref{img:compare with arwgan}.

\subsection{Experiment Setting}
All the images are resized to $128 \times 128 \times 3$, and the length of the embedded watermark message is 30. Arwgan\cite{arwgan} is chosen as our baseline network. The weights of each hyperparameter are set as s $\lambda_1$ =0.7,  $\lambda_2$=0.1, $\lambda_3$= $10^{-3}$ and $\lambda_4$=1.5, and for Adam's gradient descent, the learning rate lr =$10^{-3}$. The noise layer is designed as dropout(30\%), gaussian blur(2.0), JPEG(50), resize(0.8), random cropped(3.5\%). 

We compare the results of our method with the existing sota method in Table \ref{tb:generlization_compare} for the proposed method under various attack scenarios. Considering there are some methods that are not open source, we use the results presented in its paper for comparison and try our best to keep the same training setting. The experimental results demonstrate that our method achieves comparable and even better results under multiple datasets and multiple noisy attack scenarios.
\subsubsection{Datasets}
The proposed watermarking model is implemented by pytorch and trained on NVIDIA GeForce RTX 3090. The COCO dataset \cite{COCO} is widely used in the field of computer vision, from which we randomly select 15,000 images as our training set for the current task and 3,000 as the test set. In addition, in order to evaluate the generalization of our method, we randomly select 1000 images from voc2012\cite{pascal-voc-2012}, 1000 images from the imagenet \cite{IMageNet} dataset, and 300 images from the DIV2k \cite{div2k} dataset to test the generalization of our method across the dataset. The results are shown in the table \ref{tb:generlization_compare}.

\subsubsection{Metrics}
For digital watermarking, we typically use PSNR(Peal Signal Noisy Rate, PSNR) and SSIM(Structure Similarity Index Measure, SSIM) 
 to evaluate the visual quality of the images, higher values indicate superior visual quality. For model robustness, we use BAR(Bit Accuracy Rate, BAR) to evaluate the capability between embedded and extracted watermarks.

\subsection{Generalization Experiments}
By working on unseen datasets, we evaluate the cross-dataset generalization ability of our proposed method in comparison to existing methods. Results are shown in Table \ref{tb:generlization_compare}. The results obtained from our proposed method on several previously unseen datasets demonstrate that it is capable of achieving comparable or even superior performance to that of existing state-of-the-art methods.

\subsection{Ablation Experiments}
To assess the effectiveness of our improved component, we compared the robustness of visual quality by training different models on the training dataset. The ablation experiments intuitively show that the best robustness is obtained when using both the SeE-encoder and the denoiser module shown in table \ref{tab:ablation experiment}. 

\begin{table}[h!]
\begin{center}
\caption{Ablation Experiment Results} \label{tab:ablation experiment}
\begin{tabular}{cccccc}
  \hline
    SE-encoder& Denoise & PSNR & SSIM  & BAR  \\
\hline
 &  & 36.82 & 0.9680 & 98.78 \\
\checkmark  &  &36.79 & 0.9675 &98.83\\
 & \checkmark & \textbf{37.43} & \textbf{0.9772} & 98.92\\ 
 \checkmark & \checkmark  &37.38 & 0.9762 & \textbf{99.02}\\  
\hline
\end{tabular}
\end{center}
\end{table}

\section{Conclusion and Discuss} \label{conclusion}

 Aiming at the existing END framework algorithm's weak ability to resist noise, this paper proposes the following two improvements. (1) By introducing the SE module in the encoder to perform channel-wise and pixel-wise fusion of watermark information, the watermark features to be embedded are processed more effectively, the redundancy of watermark information is reduced and the robustness is improved. (2) The innovative introduction of the denoise module improves the robustness of the watermarking model by recovering the watermark information lost due to noise. Thus, the robustness task of watermarking can be transformed into the denoising recovery task of images to a certain extent. The denoise module proposed in this paper can be easily extended to existing watermarking networks and achieve improvements in robustness and visual quality, which we believe will suggest an improved network design scheme for future watermarking networks.

\bibliographystyle{IEEEbib}
% bibtext file name 
\bibliography{icme2023template}

% \bibitem{b1} G. Eason, B. Noble, and I. N. Sneddon, ``On certain integrals of Lipschitz-Hankel type involving products of Bessel functions,'' Phil. Trans. Roy. Soc. London, vol. A247, pp. 529--551, April 1955.
% \bibitem{b2} J. Clerk Maxwell, A Treatise on Electricity and Magnetism, 3rd ed., vol. 2. Oxford: Clarendon, 1892, pp.68--73.
% \bibitem{b3} I. S. Jacobs and C. P. Bean, ``Fine particles, thin films and exchange anisotropy,'' in Magnetism, vol. III, G. T. Rado and H. Suhl, Eds. New York: Academic, 1963, pp. 271--350.
% \bibitem{b4} K. Elissa, ``Title of paper if known,'' unpublished.
% \bibitem{b5} R. Nicole, ``Title of paper with only first word capitalized,'' J. Name Stand. Abbrev., in press.
% \bibitem{b6} Y. Yorozu, M. Hirano, K. Oka, and Y. Tagawa, ``Electron spectroscopy studies on magneto-optical media and plastic substrate interface,'' IEEE Transl. J. Magn. Japan, vol. 2, pp. 740--741, August 1987 [Digests 9th Annual Conf. Magnetics Japan, p. 301, 1982].
% \bibitem{b7} M. Young, The Technical Writer's Handbook. Mill Valley, CA: University Science, 1989.
% \end{thebibliography}
% \vspace{12pt}
% \color{red}
% IEEE conference templates contain guidance text for composing and formatting conference papers. Please ensure that all template text is removed from your conference paper prior to submission to the conference. Failure to remove the template text from your paper may result in your paper not being published.

\end{document}